# Si- and C-Induced Modifications in Vapor–Solid-Grown InGaAs Nanowires: From Crystal Structure to Carrier Dynamics

Hamidreza Esmaielpour[1,2], Leopold Rothmayer[1], Thomas Trinkl[1], Laura Niermann[2], Tore Niermann[2] Michael Lehmann[2], Jonathan J. Finley[1], and Gregor Koblmüller[1,2]

[1] Walter Schottky Institute, TUM School of Natural Sciences, Technical University of Munich, Am Coulombwall 4, 85748 Garching, Germany.

[2] Institute of Physics and Astronomy, Technical University Berlin, Hardenbergstraße 36, 10623 Berlin, Germany.

**Abstract_** Determining the impact of dopants in ternary III–V nanowires (NWs) is crucial for their applications in next-generation optoelectronic and photonic devices; however, the interplay between dopant species, crystal structure, and carrier dynamics in vapor-solid-grown InGaAs NWs remains underexplored. In this study, the influence of silicon (Si) and carbon (C) doping on the structural, morphological, and optical properties of catalyst-free, selective-area grown $In_{0.20}Ga_{0.80}As$ NWs fabricated by molecular beam epitaxy is investigated. Scanning electron microscopy demonstrates that C-doping increases the NW aspect ratio by over 100% compared to undoped NWs, while a concomitant decrease in their planar defect density is observed, indicating enhanced crystal quality with C-doping. Despite their superior crystal quality, the C-doped NWs present a diminished photoluminescence intensity, due to point defects and increased surface recombination velocity. This is further supported by time-resolved photoluminescence spectroscopy, revealing accelerated non-radiative recombination in the C-doped NWs. Lastly, hot-carrier effects in these nanostructures are studied, and their signatures remain observable, albeit with reduced intensity in both Si- and C-doped NWs. This comprehensive study establishes clear correlations between doping parameters and the resulting material characteristics and carrier dynamics, offering valuable insights into the fundamental mechanisms governing the VS-grown InGaAs NWs and their implications in optoelectronic devices.

## Introduction

III–V ternary semiconductor NWs have emerged as highly versatile building blocks for next-generation optoelectronic and photonic devices, mainly owing to their favorable one-dimensional (1D) geometry, high photoabsorptivity, and the possibility of their monolithic integration onto silicon substrates [1,2,3]. InGaAs NWs on Si are the focus of extensive research due to their potential in next-generation photovoltaics [4,5,6], integrated photonics [7,8], and high-performance gate-all-around III–V/Si transistors [9]. The small carrier effective mass, high carrier mobility, and the compositional tunability of this ternary alloy across a wide bandgap spectrum, encompassing significant near-infrared and telecommunications wavelengths, render these NWs particularly appealing for both classical and quantum photonic applications [10,11].

One promising approach to fabricating InGaAs NWs and precisely controlling their chemical composition is the catalyst-free, vapor-solid (VS) growth mode. A crucial factor influencing the performance of InGaAs NWs is their crystalline structure [12]. Unlike bulk InGaAs, which predominantly crystallizes in the zinc blende (ZB) phase [13], NWs readily exhibit a mixed polytypic phase characterized by stacking faults and rotational-twin defects [10,14]. Depending on the flux rate and growth kinetics, the NW microstructure alternates between ZB and wurtzite (WZ) domains, leading to a disordered layer stacking [15,16,17]. The fabrication of NWs by the VS-growth mode is particularly pertinent in this context, as it depends on an imbalance in facet-growth velocities and twin-defect formation, which typically yields a higher density of stacking faults than vapor–liquid–solid (VLS) growth [15,18]. Therefore, understanding and mitigating these structural imperfections are crucial for the prospective applications of VS-grown InGaAs NWs. Numerous approaches have been implemented to reduce defect density and enhance the material properties of NWs, including precise control of growth parameters [19,20], adjusting their diameters [18,21], and incorporating dilute elements or surfactants during NW fabrication [22,23,24]. Many of these approaches modify the surface energy of NW facets and promote the formation of a specific crystal phase over others, thereby reducing defect density [19,21].

An additional important aspect for NW-based devices is intentional doping, which is indispensable for controlling the carrier type, concentration, and, ultimately, the optical and transport properties [25,26,27]. Silicon (Si) and carbon (C) serve as the most prevalent n-type and p-type dopants for III–V compounds, respectively; however, owing to their amphoteric nature and growth conditions, their behavior within NWs is markedly complex [28,29]. In other words, these elements may substitute either anion or cation lattice sites contingent upon flux conditions and growth parameters, thereby partially counteracting the intended doping effects [29]. Ruhstorfer et al. [30] demonstrated that Si-doping at low fluxes in VS-grown GaAs NWs predominantly exhibits n-type characteristics, whereas increasing the doping concentration induces self-compensation effects via the formation of Si-induced complexes, leading to reduced overall n-doping properties [31]. Furthermore, doping species have the potential to alter the microstructural and optical-electrical characteristics of the system by modifying defect densities and scattering centers [32,33].

Despite significant advances in understanding the effects of individual doping elements in binary III–V NWs, a comprehensive and systematic investigation of how Si- and C-doping influences the material properties and carrier dynamics of VS-grown InGaAs NWs remains insufficiently explored. In the present study, we aim to address these aspects by providing a comprehensive structural and spectroscopic analysis of Si- and C-doped InGaAs NWs synthesized via vapor-solid molecular beam epitaxy (VS-MBE). Here, we conduct a thorough examination of the evolution of crystal phases, NW morphology, and the dynamics of carrier recombination, all as functions of the type of doping species employed and their respective concentrations.

## Results and Discussions

High-uniformity arrays of $In_{0.20}Ga_{0.80}As$ NWs were grown by catalyst-free, selective-area epitaxy on prepatterned $SiO_2$/Si(111) substrates using molecular beam epitaxy (MBE). Hereby, in the $SiO_2$-mask layer (20-nm thick) periodic mask openings (diameter of 60 nm) were created through a combination of electron beam lithography and wet chemical etching to induce site-selective NW nucleation. Details regarding the wafer preparation process and conditions for NW nucleation are presented elsewhere [30]. Five NW-arrays were grown, one intrinsically undoped, two Si-doped, and two C-doped. The NWs were grown at a substrate temperature of 595 °C, as measured by an optical pyrometer, under an arsenic ($As_4$) overpressure of $4.5 \times 10^{-5}$ mbar, utilizing a valved As cracker source. The flux rates of the group-III elements, indium (In) and gallium (Ga), were maintained at 0.31 Å/s and 0.275 Å/s, respectively, resulting in a V/III ratio of 42. The growth duration for the core InGaAs NWs was 120 min, followed by the deposition of a 49-minute long $In_{0.20}Al_{0.80}As$ shell layer aimed at passivating the NWs and enhancing their quantum yield of luminescence.

Table 1. Nominal Si- and C-doping concentrations of the InGaAs NWs determined by secondary ion mass spectrometry on GaAs(001) planar reference layers..

| Dopants | Heating current (A) | Doping Concentration ($cm^{-3}$) |
|---|---|---|
| Si | 8 | $[n] = 1 \times 10^{18}$ |
| | 9 | $[n] = 3 \times 10^{18}$ |
| C | 46 | $[p] = 4 \times 10^{17}$ |
| | 49 | $[p] = 2 \times 10^{18}$ |

Doping was performed only during the core-InGaAs NW growth, while the InAlAs shell remains undoped in all structures. In this study, Si- and C-dopants were introduced by heating a silicon filament and a graphite filament within sublimation sources during NW growth. The evaporation rates of these dopants are controlled through precise adjustments to the heating current (in units of Ampere) supplied to the filaments, and calibrated using planar GaAs reference layers. Table 1 provides information regarding the nominal Si- and C-concentrations as determined by Secondary Ion Mass Spectrometry (SIMS) on the GaAs(001) planar reference layers. Although the chemical dopant concentration was not measured directly in the NW samples, we anticipate similar concentration values as for the planar reference layers. This was recently confirmed in correlated atom probe tomography studies performed on Si-doped GaAs NWs [31]. Likewise, unintentionally doped InGaAs NWs in the Ga-rich limit exhibit intrinsically n-type behavior with an estimated doping concentration in the $\sim 10^{17}$ $cm^{-3}$ range, based on recent studies [10,34,35].

Figure 1 shows scanning electron microscopy (SEM) images of the InGaAs NWs: (a) undoped, (b) low Si-doped (heating current of 8A), and (c) low C-doped (heating current of 46A). High-resolution X-ray diffraction (XRD) measurements confirm identical chemical compositions across all the investigated samples, irrespective of doping. It is observed that the Si-doped NWs exhibit a larger diameter compared to their undoped counterpart, while their length is reduced(see also Figure S1 in the Supporting Material). Conversely, the C-doped NWs demonstrate higher axial growth rates than the undoped NWs. This effect is reflected in the aspect ratio (the ratio of length to diameter) of the NWs, wherein C(46A)-doping approximately doubles the aspect ratio compared to undoped NWs, whereas Si-doping markedly decreases the aspect ratio of the InGaAs NWs. Salehzadeh et al. [36] observed a similar enhancement in the axial growth rate of gold-assisted GaAs NWs grown in the presence of $CBr_4$ precursor, and attributed this striking effect to the suppression of 2D planar growth induced by this precursor. Also, a possible passivation of dangling bonds on the surface of the GaAs NWs by the precursor was proposed [36], leading to enhanced adatom surface diffusion and an increased axial growth rate.

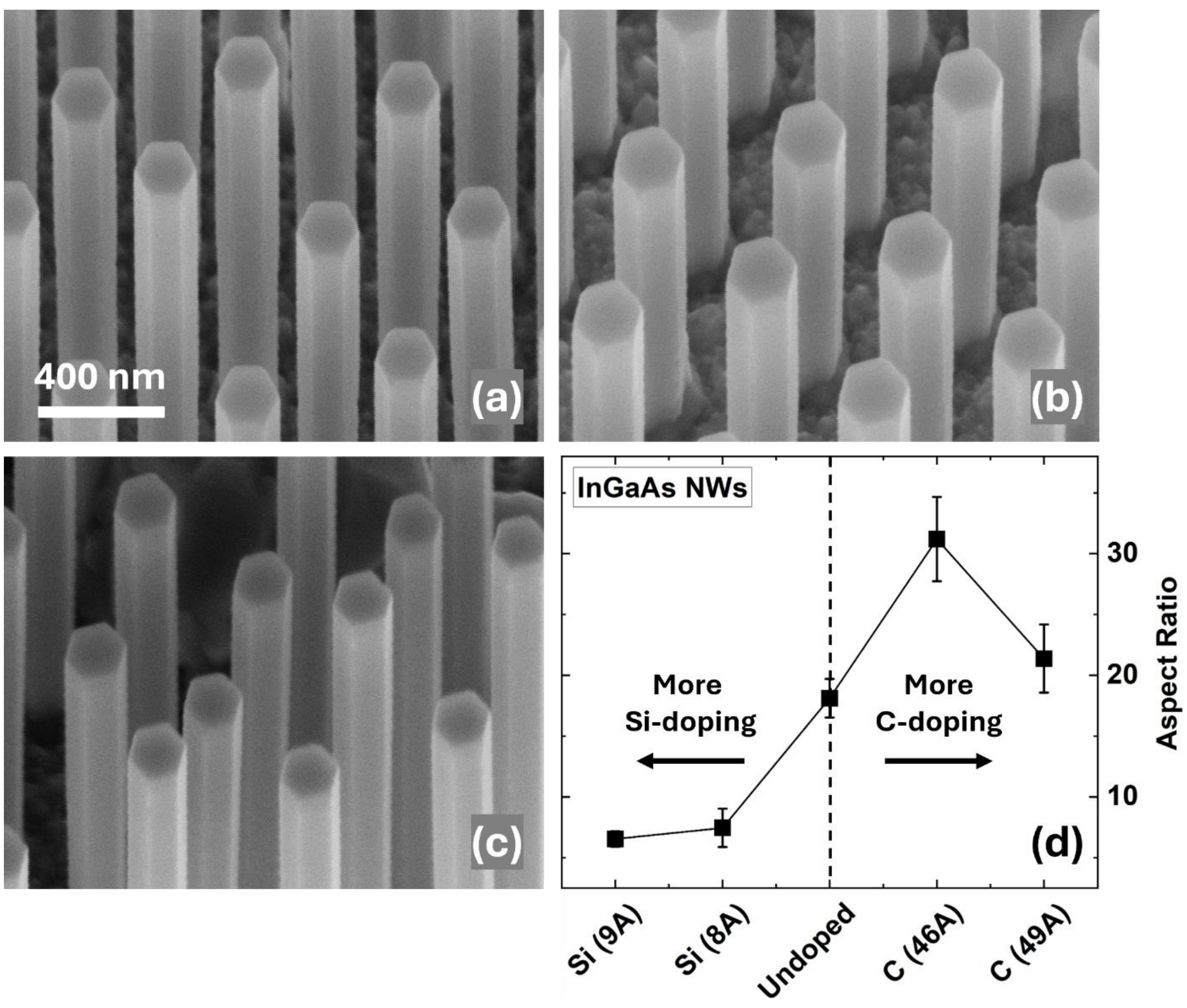

Figure 1. SEM images of the (a) undoped, (b) Si(8A)-doped, and (c) C(46A)-doped InGaAs NW arrays. (d) Aspect ratio of the InGaAs NWs of different doping species and concentrations.

The reduction in axial growth rate by Si-doping was also observed in Si-doped GaAs(Sb) NWs [23] and InAs NWs [37], where the doped NWs exhibit a decrease in their length. The underlying cause of this enhanced lateral growth was attributed to Si segregation at the NW sidewall and consequent reduction in the adatom diffusion length of group-III species, promoting radial growth along the side facets [37].

The doping species further influence the microstructural properties of the NW, such as crystal layer stacking and defect density. To obtain deeper insight into the distinct growth dynamics of Si- and C- doping in the InGaAs NWs, high-resolution scanning transmission-electron microscopy (HR-STEM) was conducted, as illustrated in Figure 2. The high angle annular dark field (HAADF) images in Figure 2(a) and (b) show small segments of the Si- and C-doped NWs, respectively; however, the analysis to estimate the defect density, as shown in Figure 2(c), was performed for various sections along the NWs to provide statistically relevant data. The STEM results reveal that while the ZB phase remains the predominant crystal structure, rotational twins are observed along the NWs. Such planar defects generate atomically sharp local type-II band alignment, which can trap carriers and create scattering centers, thereby degrading the optical and electrical performance of NWs [12,38]. Figure 2(a) demonstrates that the incorporation of Si into the core NW substantially increases the density of twin defects along the growth direction, consequently affecting the crystal phase purity. Similar observations of increased defect density due to Si-doping have been reported in various III-V NWs [21,39].

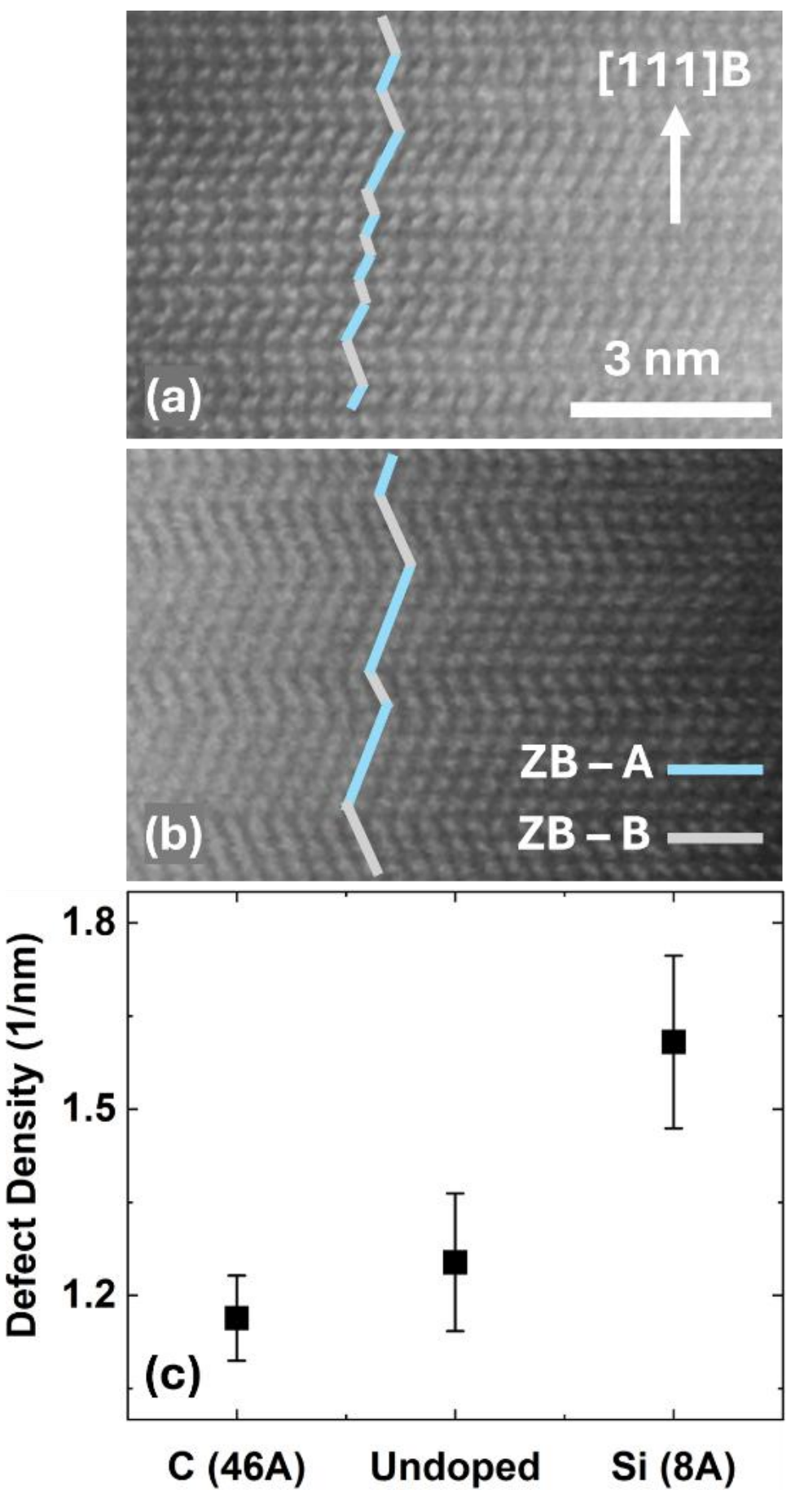

Figure 2. High-resolution TEM micrographs of the (a) Si(8A)-doped and (b) C(46A)-doped InGaAs NWs. Zincblende (ZB) crystal layer stacking is observed, but with a high density of rotational twin domains (labeled as ZB-A, ZB-B). (c) Density of rotational twin defects for the InGaAs NWs of different doping. The change in contrast and brightness is due to a thickness gradient in that region.

In contrast, C-doping demonstrates an enhancement in crystal quality by decreasing the density of twin defects, as illustrated in Figure 2(c), leading to larger ZB domains within the InGaAs NWs.

A comparable observation has been reported in VLS-grown C-doped GaAs NWs, whereby in this growth mode a strong tendency towards phase-pure ZB crystal structure devoid of planar defects is seen [36]. The origin of such an enhanced crystal quality was ascribed to an increased axial growth rate, which leads to a decreased proportion of WZ to ZB. Similar findings regarding the relationship between the axial growth rate and the phase purity of NWs have also been reported by Joyce et al. [20] and Spirkoska et al. [40] concerning VLS-grown GaAs NWs.

Furthermore, dopant species can alter the optical properties of nanostructures by introducing sub-gap states and modifications in the crystal lattice [33]. Figure 3 illustrates the PL spectra of (a) the undoped, (b) the Si(8A)-doped, and (c) the C(46A)-doped InGaAs NWs at 10 K under various excitation powers, revealing discernible differences among the NWs. The PL spectra of the Si(9A)-doped and C(49A)-doped InGaAs NWs at 10 K are shown in Figure S2 in the Supporting Material for reference. The findings demonstrate that doping leads to linewidth broadening in the PL spectra and clearly affects the emission intensity. The increase in spectral linewidth is more pronounced in the Si-doped NWs than in the C-doped ones, due to greater crystalline disorder resulting from dopant incorporation and the band-filling (or Burstein-Moss) effect [33]. Arab et al. [35] examined the PL emission of Si-doped GaAs NWs and observed an increase in linewidth broadening correlated with increasing doping concentrations. They attributed this spectral broadening to the formation of various recombination centers induced by the doping process. In a separate study conducted by Ruhstorfer et al. [30], the impact of Si doping on the optical properties of GaAs NWs was also investigated, with findings indicating that increased doping concentrations lead to an expansion of the PL linewidth and a blue-shift of the peak position towards higher energies, attributable to the band-filling effect.

Furthermore, the C-doped NWs also exhibit an increase in linewidth broadening relative to the undoped NWs. The origin of this effect is attributed to the Burstein-Moss effect, arising from a combination of direct and indirect optical transitions driven by deeper Fermi-level penetration into the valence band [41,42]. As the excitation power increases, optical transitions are no longer confined to the band edge but can originate from a broader range of k-states, thereby smearing the emission spectrum [43]. Additionally, the relaxation of k-selection rules, induced by the disruption of crystal periodicity, makes phonon-assisted and indirect transitions optically active, thereby increasing the spectral width [41,44,45].

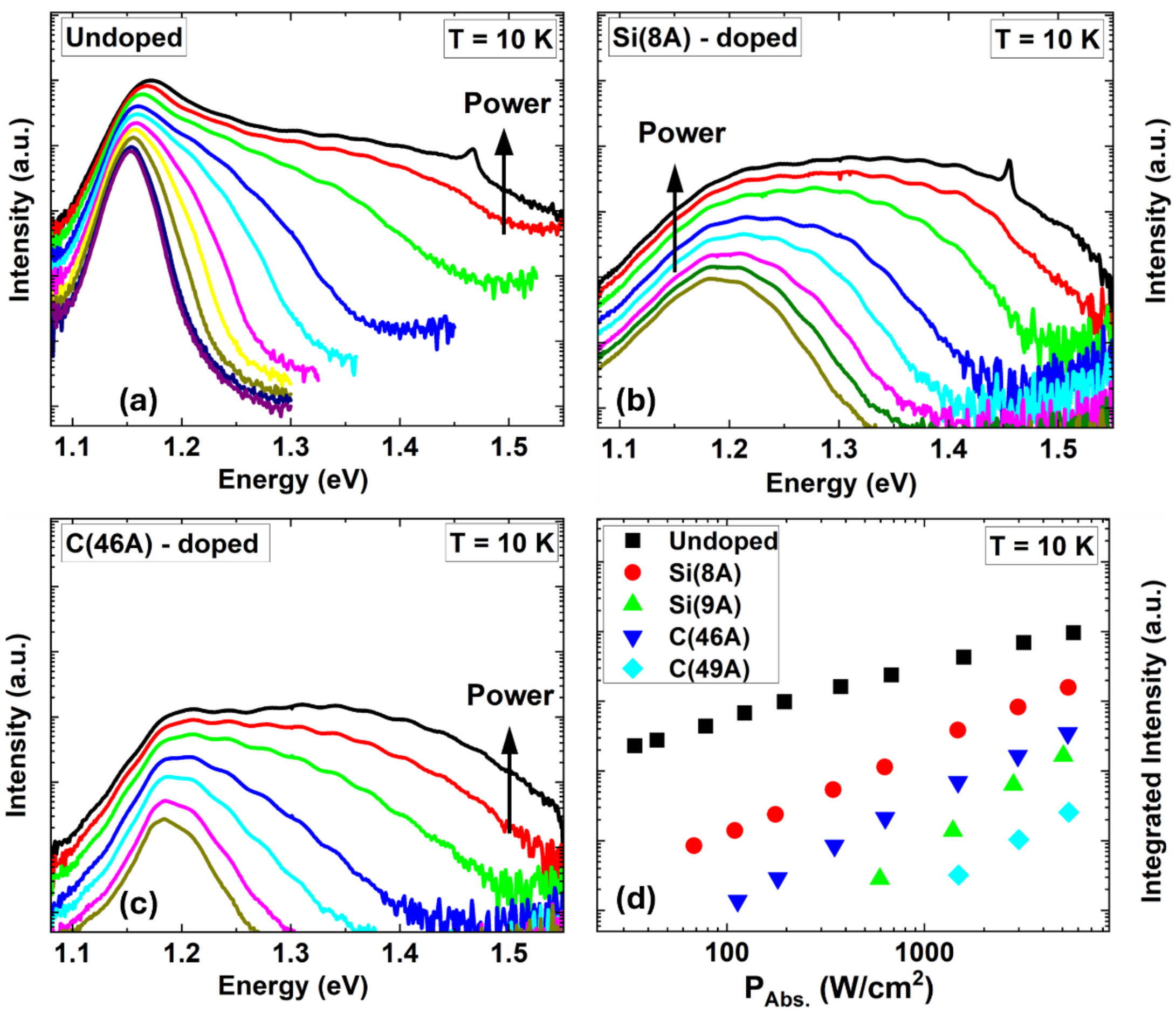

Figure 3. PL spectra of the (a) undoped, (b) Si(8A)-doped, and (c) C(46A)-doped InGaAs NWs at 10 K under absorbed power densities from 30 $W/cm^2$ to 6000 $W/cm^2$. (d) Power-dependent integrated PL intensity of the NWs of varying doping species and concentrations at 10 K.

In addition, the influence of doping on the NWs is also reflected in their quantum yield of luminescence due to modifications of carrier dynamics at the presence of defects [42]. Figure 3(d) illustrates the results of the integrated PL intensity of the NWs at 10 K versus the absorbed power density. Increasing the excitation power increases the emitted PL intensity, irrespective of doping. However, while the undoped NWs maintain the highest intensity, the Si- and C-doped samples show an up to three orders of magnitude decrease in intensity. The stronger decrease in the quantum yield of luminescence observed in the C-doped InGaAs NWs compared with the Si-doped NWs (despite their lower dopant concentration), is most likely arising from an interplay between minority-carrier dynamics, surface recombination, and carbon-induced defects [46,47,48]. The rate equation analysis is carried out for the NWs with various dopant species, as illustrated in Figure S3 in the Supporting Material. The results indicate that, while Auger recombination predominantly contributes to the undoped NWs, increasing doping concentrations results in a shift in recombination dynamics toward the Shockley-Read-Hall (SRH) mechanism. In the Si- and C-doped NWs, one main distinction emerges from the nature of the minority carriers. In the C-doped NWs, the minority carriers are electrons, which exhibit substantially higher mobilities than holes in GaAs, due to the mismatch in their mass, by approximately an order of magnitude [49]. This elevated mobility enables their diffusion toward NW facets, where they can be captured by surface trap states and undergo non-radiative recombination. Conversely, in the Si-doped NWs, holes serve as the minority carriers, and due to their comparatively lower mobility, they are more prone to recombine within the bulk prior to reaching the surface, thereby increasing the likelihood of radiative recombination [46]. Consequently, the inherent differences in the nature of the minority carriers in these doped structures significantly influence their recombination dynamics.

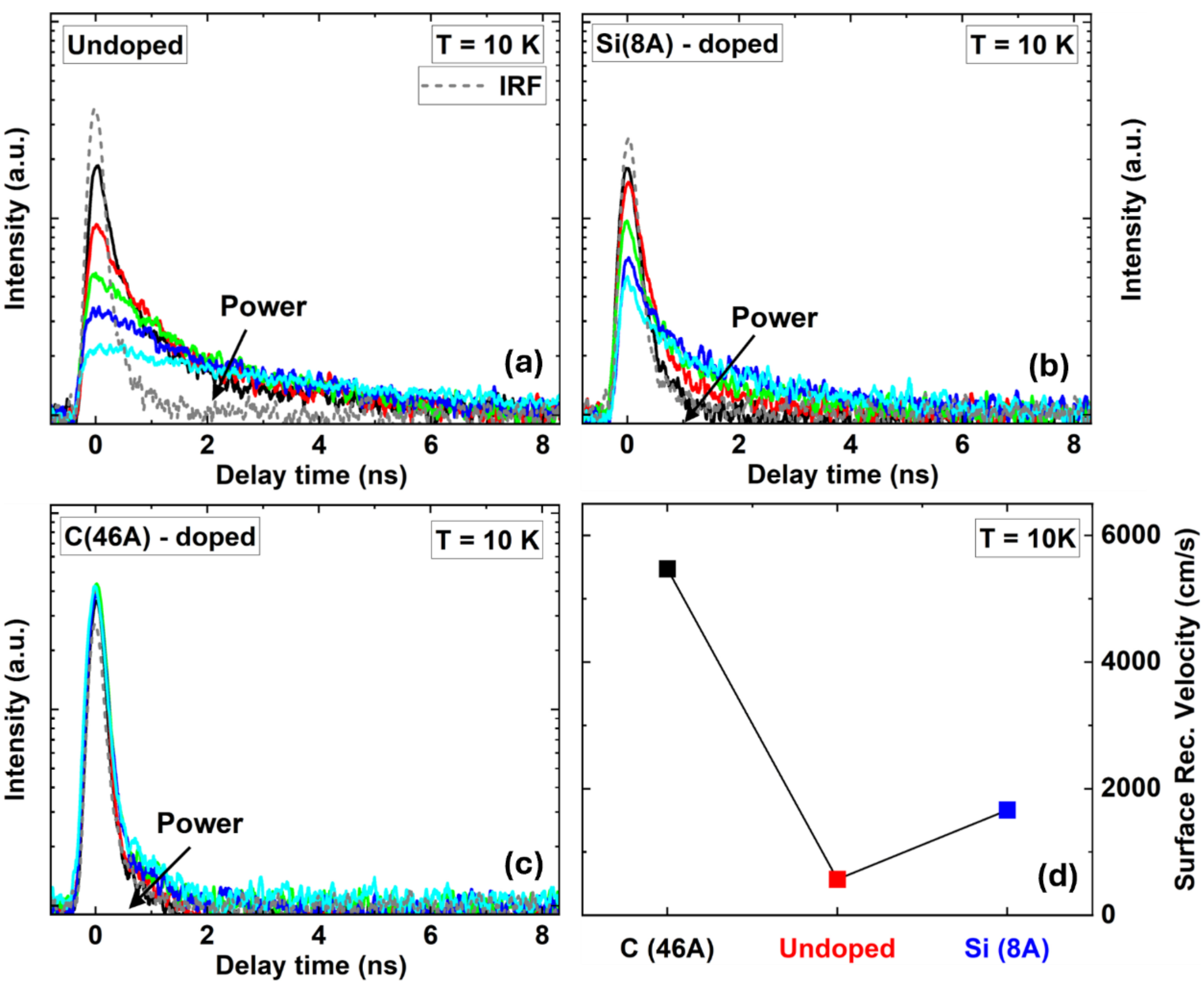

Figure 4. TRPL spectra at 10 K of (a) the undoped, (b) Si(8A)-doped, and (c) C(46A)-doped InGaAs NWs at various absorbed powers densities from 30 W/cm$^2$ to 6000 W/cm$^2$. The dashed-gray line indicates the instrument response function (IRF) of the optical setup. (d) Surface recombination velocity for different NWs at 10 K. It is seen that the undoped and the C-doped NWs show the lowest and the highest recombination velocities, respectively.

Furthermore, Boland et al. [49] studied the impact of band bending on the surface of Si- and C-doped GaAs NWs on the recombination dynamics of charge carriers. Their findings indicate that the downward band bending, induced by C-doping on the surface of the GaAs NWs [50], facilitates the injection of electrons into surface traps, thereby accelerating their non-radiative recombination [49]. In contrast, the n-doped NWs exhibit upward band-bending at the surface, and since holes, as minority carriers in these nanostructures, have lower mobility than electrons, they are therefore more likely to recombine within the bulk before reaching the nonradiative surface states. This phenomenon can indeed be effectively examined by determining the surface recombination velocity in the InGaAs NWs using time-resolved photoluminescence (TRPL) spectroscopy [12,51].

Figure 4 illustrates the TRPL results of the InGaAs NWs under various excitation power densities at 10 K, and the dashed-gray line shows the instrument response function (IRF) of the system. The results of TRPL signals at 10 K for the Si(9A)- and C(49A)-doped NWs are shown in Figure S4 in the Supporting Materials. The results indicate a strong dependence in the recombination dynamics of photogenerated carriers on dopants in the NWs. At higher excitation power densities, where the contributions of Auger and radiative recombination are large, the decay rate is higher, whereas reducing the power density can intensify the contribution of the Shockley-Read-Hall (SRH) process in the system [52]. Moreover, the comparison of the TRPL results for the (a) undoped and the (b) Si- and (c) C-doped NWs indicates that the recombination dynamics accelerate when the InGaAs NWs are doped with Si and C, consistent with the results of the integrated PL intensity of the NWs, which originates from accelerated non-radiative recombination. Notably, this effect is significant for the C-doped NWs, particularly at higher excitation powers, where the accelerated recombination is beyond the time resolution of the optical apparatus. To determine the influence of trap states at the surface on the dynamics of recombination, the surface recombination velocity ($v_s$) of the NWs is determined using the following equation [51]:

$$\tau_{TRPL}{}^{-1} = \left(\frac{4}{d}\right) v_s + \tau_{Bulk}{}^{-1}, \tag{1}$$

where "$\tau_{TRPL}$" and "$\tau_{Bulk}$" are the lifetimes determined from the TRPL spectra and from the bulk, respectively, and "$d$" is the NW diameter. The bulk lifetime is estimated to be 18 ns at 10 K, determined from core-shell InGaAs NWs of similar chemical composition [12], which is consistent with reported values for the recombination lifetime of core-shell InGaAs/InP NWs [53]. Figure 4(d) illustrates the surface recombination velocity for the undoped and doped NWs, and it is seen that the undoped NWs possess the lowest surface recombination velocity, which agrees with their higher PL intensities. Additionally, the results indicate that the C(46A)-doped NWs have the highest surface recombination velocity, suggesting that their higher rates of non-radiative recombination, despite their improved crystal quality, could originate from enhanced recombination at the surface [49].

In addition, the recombination dynamics can also influence the properties of non-equilibrium photogenerated carriers and their rates of thermalization in the system. To investigate these dynamics, the emitted PL spectra are analyzed by the generalized Planck's radiation law, as given by [54,55]:

$$I_{PL}(E) = \frac{2\pi\, A(E)\, (E)^2}{h^3 c^2} \left[ exp\left(\frac{E - \Delta\mu}{k_B T}\right) - 1 \right]^{-1}, \tag{2}$$

where, "$I_{PL}$" denotes the emitted PL intensity, "$A(E)$" the energy-dependent absorptivity, "$h$" Planck's constant, "$k_B$" Boltzmann's constant, and "$c$" the speed of light. The thermodynamic properties of hot carriers are characterized by the carrier temperature "$T$" and the quasi-Fermi level splitting "$\Delta\mu$", respectively. The details of the analysis for similar material systems are presented elsewhere [6,56].

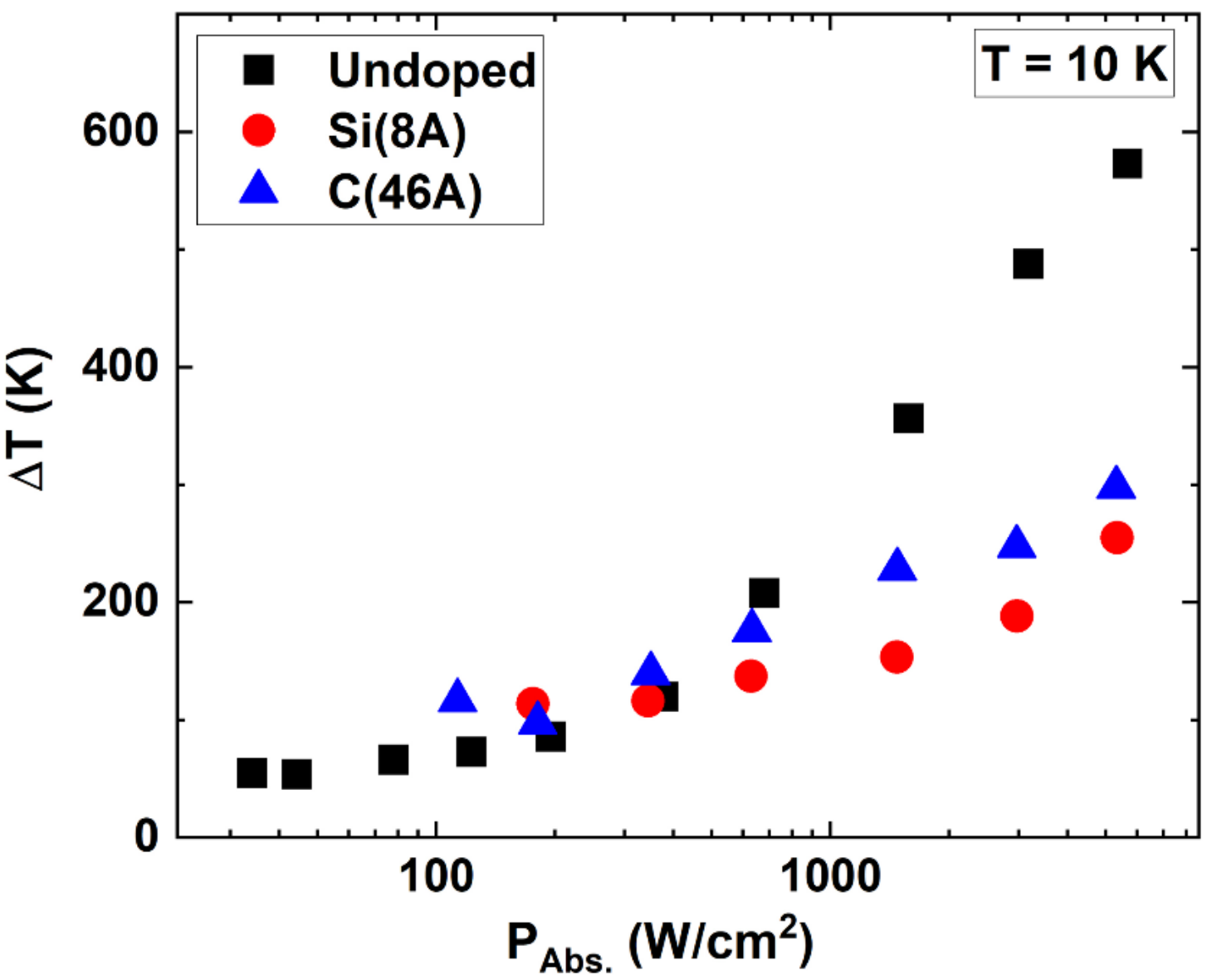

Figure 5. Hot-carrier temperature of the InGaAs NWs with different doping as a function of the absorbed power density at 10 K.

Figure 5 demonstrates the dependence of the non-equilibrium hot carrier temperature ($\Delta T$: the temperature difference between the hot carrier temperature and the lattice) on the absorbed power density at 10 K for the three NW arrays. The system without doping has the highest carrier temperature, and the effects of hot carriers diminish as the NWs are doped with Si and C. The weaker hot-carrier effects in doped NWs are attributed to increased non-radiative thermalization channels, which accelerate hot-carrier relaxation and lower their temperature. These results are significant for NW-based hot-carrier solar cells, where the contributions of hot carriers in doped regions can further improve the performance of these devices.

## Conclusions

In summary, we have presented a thorough study of the effects of Si- and C-doping on the structural, morphological, and optical characteristics of catalyst-free, VS-grown InGaAs NWs fabricated by selective-area molecular beam epitaxy. The results indicate that the C-doped InGaAs NWs demonstrate a significant increase in axial growth rate and improved crystal quality with a reduced density of planar defects. Additionally, upon Si- and C-doping, the PL spectra broaden, with the effect being more pronounced in Si-doped NWs, consistent with their higher defect density and the band-filling effect. Despite the higher crystal quality of C-doped NWs, their integrated PL intensity diminishes, and their TRPL signals indicate an enhancement in the decay rates of photogenerated carriers within these nanostructures. The findings indicate that the origin of such accelerated non-radiative recombination is attributed to the enhanced surface recombination velocity, which can deplete carriers non-radiatively. Ultimately, evidence of non-equilibrium hot populations is observed in the InGaAs NWs (undoped and doped), a phenomenon of significance for hot-carrier-based devices.

## Acknowledgements

This work was supported by the Deutsche Forschungsgemeinschaft (German Research Foundation, DFG) through Germany's Excellence Strategy via the Cluster of Excellence e-conversion (EXC 2089/1-390776260). HE also acknowledges Marie Sklodowska-Curie Actions (MSCA) for support via the TUM EuroTechPostdoc2 Grant Agreement (No. 899987). LN acknowledges funding by the Deutsche Forschungsgemeinschaft via the Emmy Noether Programme – Project No. 578097784.

# Si- and C-Induced Modifications in Vapor–Solid-Grown InGaAs Nanowires: From Crystal Structure to Carrier Dynamics

Hamidreza Esmaielpour[1,2], Leopold Rothmayer[1], Thomas Trinkl[1], Laura Niermann[2], Tore Niermann[2]
Michael Lehmann[2], Jonathan J. Finley[1], and Gregor Koblmüller[1,2]

[1] Walter Schottky Institute, TUM School of Natural Sciences, Technical University of Munich, Am Coulombwall 4, 85748 Garching, Germany.

[2] Institute of Physics and Astronomy, Technical University Berlin, Hardenbergstrasse 36, 10623 Berlin, Germany.

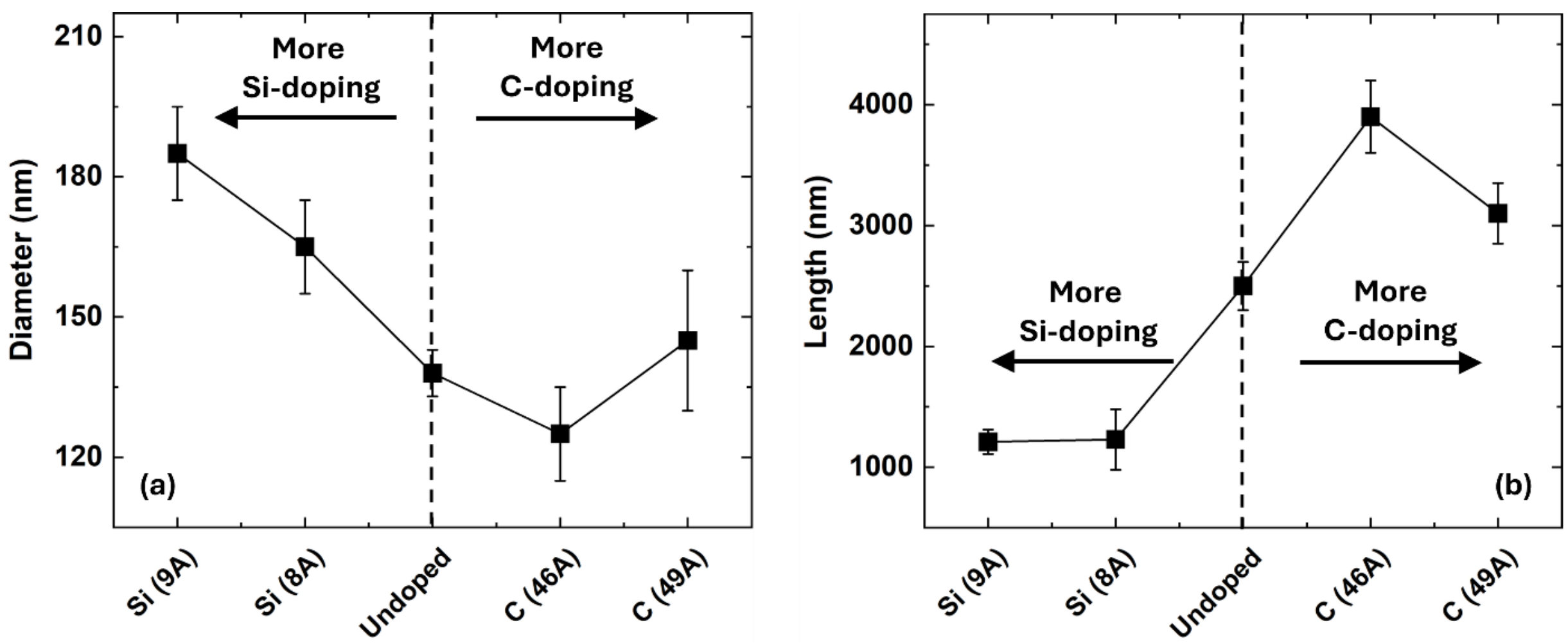

Figure S1. Dependence of the diameter and length of InGaAs NWs with various doping species and concentrations is observed. It is noted that increasing Si doping results in thicker NWs compared to undoped structures, while their length diminishes. Conversely, C-doped NWs with a 46A heating current exhibit the smallest diameter and the greatest length. However, at the highest level of C doping, (49A), the radial growth rate increases, whereas the axial growth rate decreases to lower values.

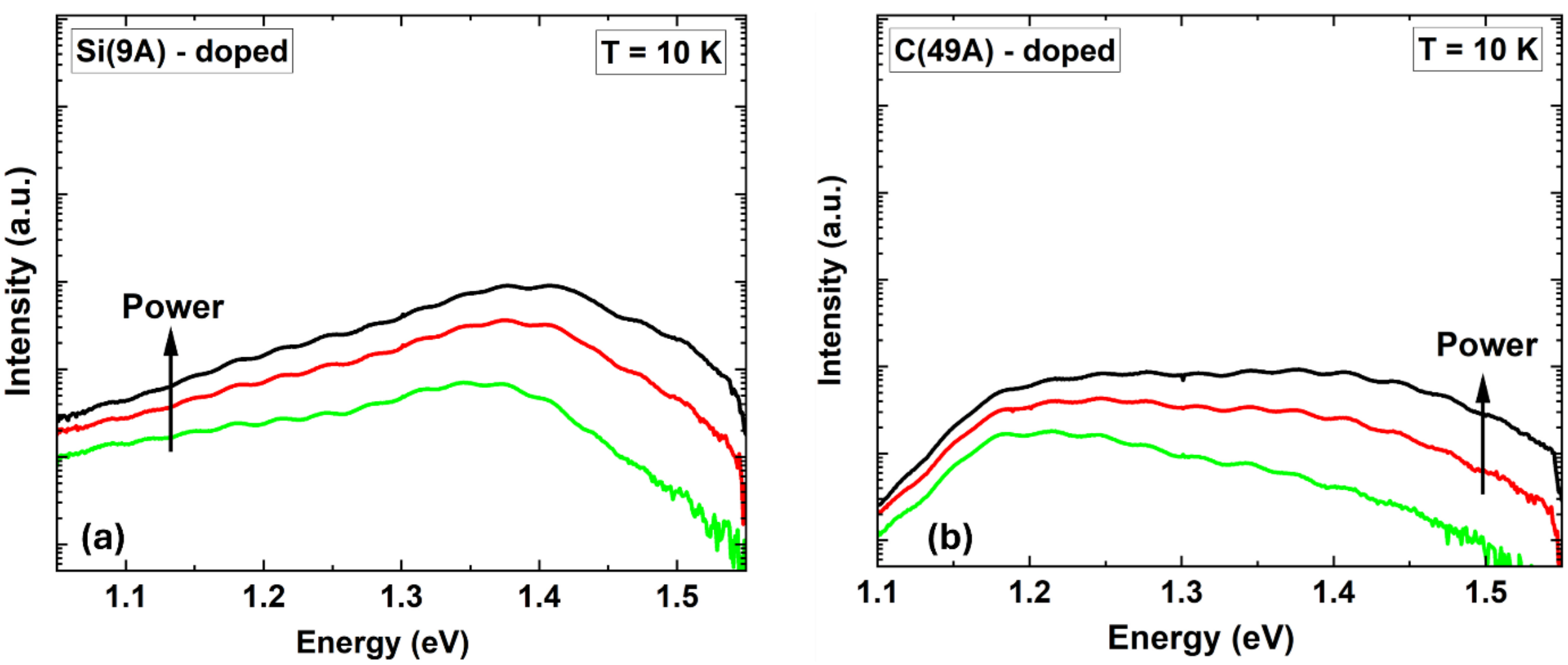

Figure S2. The PL spectra of the Si(9A)- and C(49A)-doped InGaAs NWs at 10 K under various excitation powers.

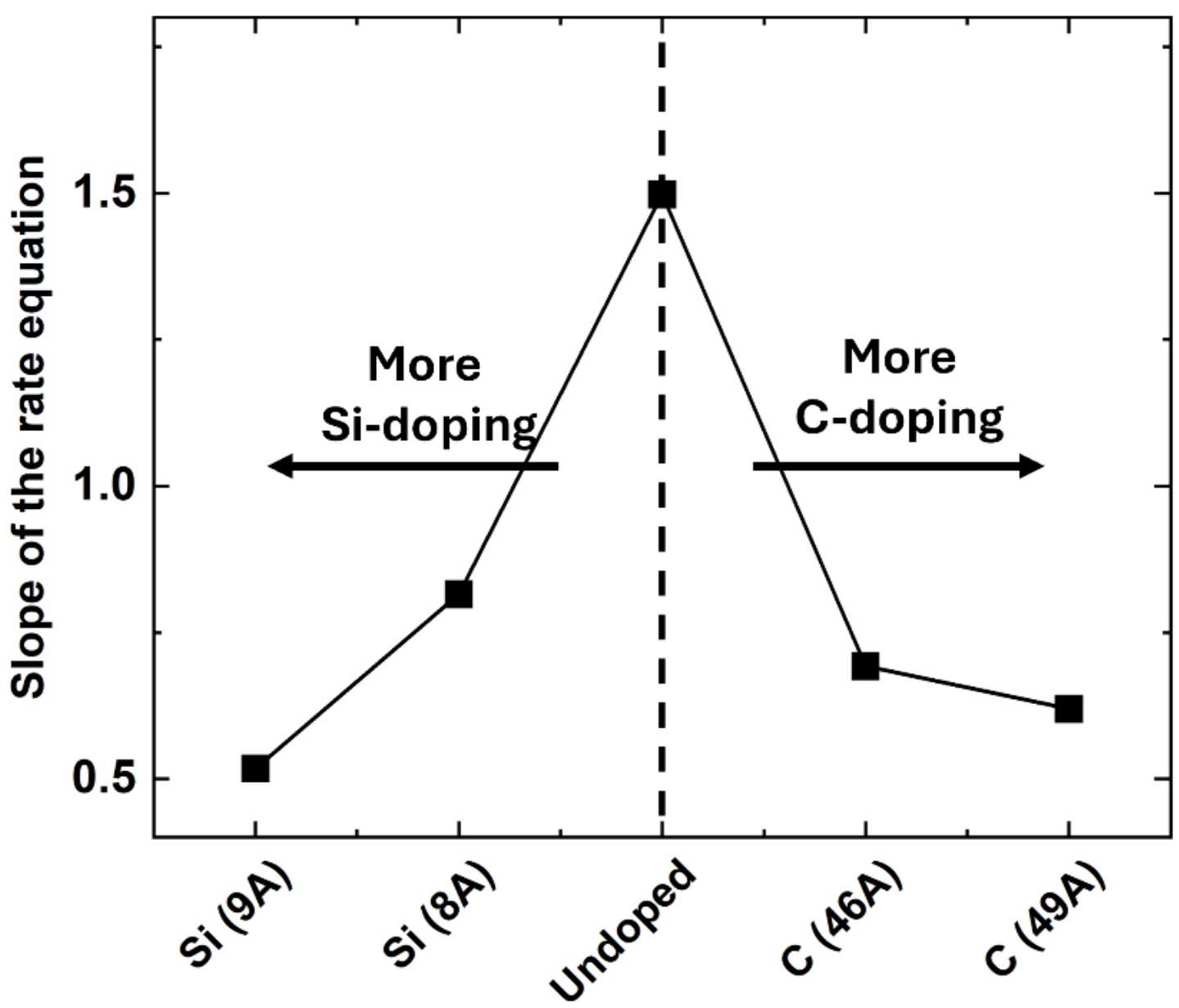

Figure S3. The results of rate-equation analysis for InGaAs NWs with various dopings. The rate equation model is given by:

$$P_{Abs.} = A\, I_{PL}^{1/2} + B\, I_{PL} + C\ I_{PL}^{3/2}. \tag{S1}$$

The first term corresponds to Shockley-Read-Hall (SRH) recombination, the second to radiative recombination, and the third to Auger recombination. The coefficients assigned to these mechanisms are A, B, and C, respectively. Their relative impacts can be directly determined from the slope of the absorbed power density versus the integrated PL intensity in a double logarithmic plot. It is seen that Auger recombination is strongest for the undoped NWs, i.e. the slope is ~1.5, while by increasing doping (both Si- and C-doped), the recombination dynamics shift toward SRH recombination.

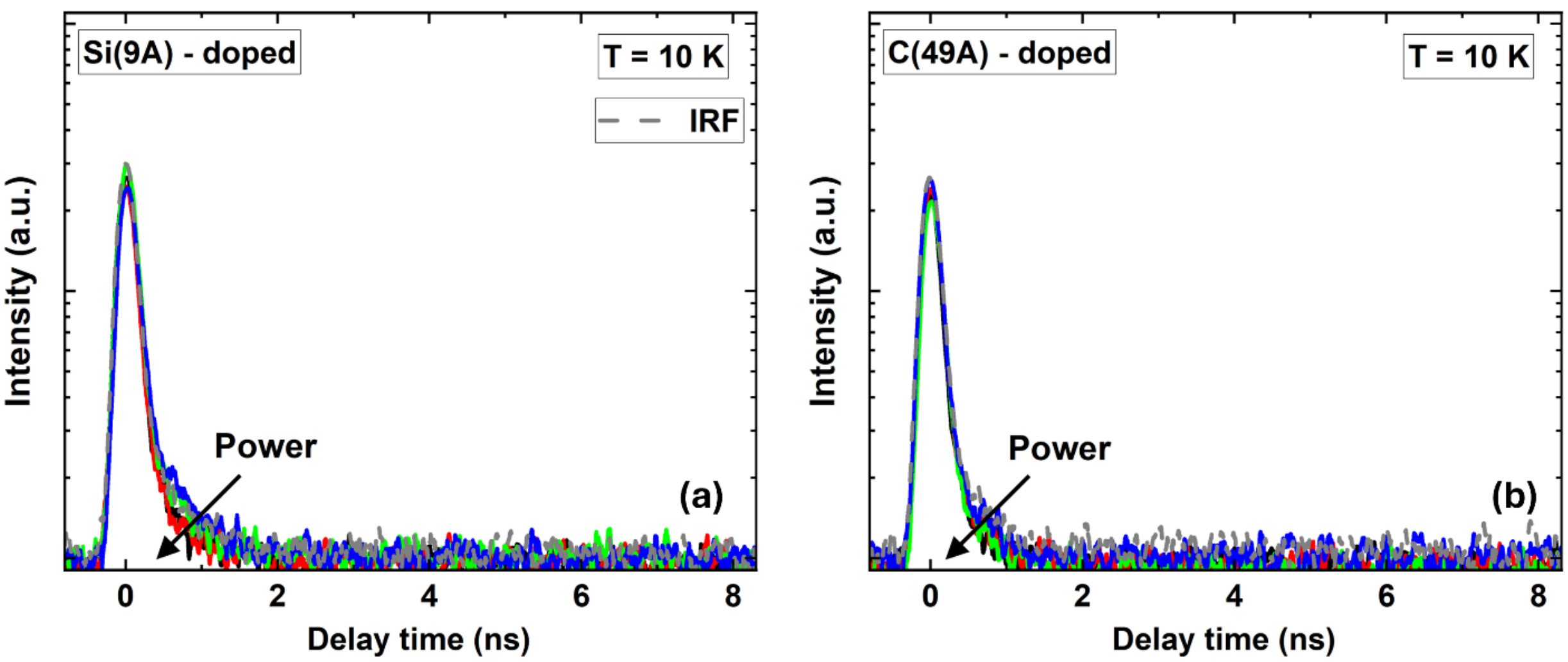

Figure S4. TRPL signals of the Si(9A)- and C(49A)-doped NWs at 10 K under various excitation powers. No noticeable change in the spectra is observed, as they exhibit behavior analogous to the instrument response function (IRF), suggesting that the recombination dynamics occur at a rate faster than the temporal resolution of the TRPL apparatus.